\documentstyle[12pt,my_ref]{article} %mine for refs - change back %once bbl included
\def\eg{{\it e.g.\ }}
\def\ie{{\it i.e.\ }}

\def\etc{{\it etc.\ }}

\def\ch2{$\chi^2$}

\def\scm  {$\hbox{{\rm cm}}^{-2}$}    %cm-2
\def\rr {\rightarrow}

\def \HI {H{\sc \,i}}

\def\lapp{\ifmmode\stackrel{<}{_{\sim}}\else$\stackrel{<}{_{\sim}}$\fi}
\def\gapp{\ifmmode\stackrel{>}{_{\sim}}\else$\stackrel{>}{_{\sim}}$\fi}

\def\@bcite#1#2{({#1\if@tempswa , #2\fi})}
\def\@pcite#1#2{#1\if@tempswa , #2\fi} %old way of writing citations

\begin{document}
%
% Title
% Capitalise the title normally - do not use ALL CAPS.
%
\title{A Catalogue of Damped Lyman Alpha Absorption Systems and Radio Flux Densities of the
Background Quasars}
%

% Authors
% Here comes the author(s) of the paper. Please add the appropriate author
% names for your paper and indicate within the $^...$ the number(s)
% which corresponds to the institute(s) of each author. In this example
% the second author has two institutional affiliations.
% Add or remove authors as required.
% **** IMPORTANT: Leave the closing curly bracket line as is. ******

\author{S. J. Curran$^{1}$, 
 J. K. Webb$^{1}$,
 M. T. Murphy$^{1}$, 
R. Bandiera$^{2}$,\\
E. Corbelli$^{2}$  and V. V. Flambaum$^{1}$
} % IMPORTANT: leave this curly bracket as the first character of this line.

% Date - leave this blank.
\date{}
\maketitle

% Institutions
% Here fill in your institute name(s) and address(es)
% The number in $^...$ indicates the author number.  For example
{\center
$^1$ School of Physics, University of New South Wales, Sydney NSW 2052, Australia, sjc@bat.phys.unsw.edu.au\\[3mm]
$^2$ Osservatorio Astrofisico di Arcetri, Largo E. Fermi, 5, 50125, Italy\\[3mm]
}

% Abstract
% Simply place your abstract between the \begin{abstract} and
% \end{abstract} commands.
%
\begin{abstract}
% Place the abstract here.
We present a catalogue of the 322 damped Lyman alpha absorbers taken
from the literature.  All damped Lyman alpha absorbers are included,
with no selection on redshift or quasar magnitude.  Of these, 123 are
candidates and await confirmation using high resolution
spectroscopy. For all 322 objects we catalogue the radio properties of
the background quasars, where known.  Around 60 quasars have radio
flux densities above 0.1 Jy and approximately half of these have
optical magnitudes brighter than $V = 18$.  This compilation should
prove useful in several areas of extragalactic/cosmological research.

\end{abstract}

cosmology: early universe
% Place keywords here. Please write all keywords in lower case. PASA uses the
 %standard list of subject 
% headings adopted by The Astrophysical Journal and available from URL:
%   http://www.journals.uchicago.edu/ApJ/keywords_text.html

% A formatting command to add space between the author list and the body
% of the paper when printed. This spacing may be changed as desired.
\bigskip

%
% Body of paper
%

\section{Introduction}

High resolution spectroscopy of quasars reveals large numbers of
absorption lines due to neutral atomic hydrogen.  In some cases, the
neutral hydrogen column density can be very large ($N_{\rm
HI}\gapp10^{20}$ \scm)\footnote{In the catalogue some column densities
less than this are to be found. This is because in order to produce
the most comprehensive list, we have also included any Lyman alpha
absorbers which are designated as possibly damped in the literature
(see Section 2).} and gives rise to a heavily damped absorption feature
(\eg \pcite{wtsc86}). The aim of this paper is to present a catalogue
of these damped Lyman alpha absorption systems (DLAs) and, where
known, the radio flux densities of the background quasar. The catalogue has
been compiled from the large number of studies reported in the
literature.  

The nature of DLAs is still open to debate. Interpretations include
galactic disks \cite{wtsc86}, low surface brightness galaxies \cite{jbm99}
or dwarf galaxies \cite{mmv97} intersecting the sight-line towards
a background quasar. Despite these possible differing morphologies, DLAs
provide a powerful cosmological probe. The highest redshift ($ z>
3.5$) DLAs may account for a large fraction of the baryons at high
redshift, suggesting they reveal gas prior to the bulk of the star
formation history of the universe (\pcite{pmsi01} and references
therein). On the other hand, recent work \cite{lyp+02} seems to
indicate star formation rates which continue to increase with
increasing redshift up to the highest galaxy redshifts observed in the
Hubble Deep Field. The discovery of a DLA with truly primordial
abundances would have a major impact on our understanding of the early
chemical evolution of the universe, and a crucial reality check on the
ever-elusive population III. This will also be important for studies
of primordial deuterium abundances (see below), since deuterium is
destroyed, and never created, by star formation and evolution.

High resolution spectroscopy can be used to study high chemical
abundances over a large redshift range. In particular, the difficult
ionisation corrections required to derive meaningful chemical
abundances in Lyman-limit absorbers (where $\log N_{\rm HI} > 17.2$
\scm, so that they are optically thick to Lyman continuum radiation,
\eg \pcite{lan91}) can be avoided using DLAs since the observed
hydrogen is probably all neutral \cite{twl+89,lwt+91}. Additionally,
at high neutral hydrogen column densities, species such as Zn{\sc \,ii}
and Cr{\sc \,ii} may become detectable, which are important since
depletion onto dust grains is thought to be negligible for the former,
whereas the latter remains in the solid phase. This allows both the
study of abundances and depletion patterns/dust reddening (see
\pcite{pbh90,pksh97}). Some further reasons why DLAs are of interest are:
\begin{enumerate}
	\item Studies of the higher order hydrogen Lyman series in
DLAs can be used to investigate the primordial deuterium abundance
\cite{wcip91}.  The advantage of using DLAs is that the deuterium
column density can be somewhat larger than typical Lyman forest
absorbers.  This may help to discriminate against H{\sc \,i}
interlopers mimicking the deuterium line.  Two recent observational
studies \cite{ddm01,pb01} report such D/H measurements.
\item Radio observations of quasars with a sufficiently high radio flux density can
provide information complementary to that of the DLA observations:
21cm H{\sc \,i} measurements reveal more
detailed kinematic information since line saturation is less severe
and provide a direct spin temperature of the cool component of the
gas. Different radio and optical morphology of the background quasar
also provides the opportunity of observing along slightly different
sight-lines through the same absorption complex, with the potential of
learning about the relative sizes of optical/radio emission regions
and the cloud size of the absorbing gas. In those rare cases where the host
quasar has a sufficiently strong millimetre flux and a foreground
molecular cloud occults the quasar \cite{wc94,wc96}, a wealth of detailed
chemistry is revealed \cite{gpb+97,cw99}.
	\item Studies of high redshift dust in DLAs gives a handle on
	the chemical evolution and star formation rates at various
	cosmological epochs \cite{pfh99} through the contribution of dust
to quasar spectral energy distributions (\eg \pcite{kvg+96,bcm+00,cbm+00,ocb+01}).
\item Certain heavy element transitions provide cosmological probes of
special interest. For example, species with ground and excited state
transitions sufficiently close to each other in energy provide a unique means of measuring the cosmic
microwave background temperature at high redshift \cite{bjl73,mbc+86,spl00}.
\item Finally, recent detailed studies of the relative positions of
heavy element atomic optical transitions and comparison with present
day (laboratory) wavelengths, suggests that the fine-structure
constant ($\alpha\equiv e^2/\hbar c$) may have evolved with
time. Inter-comparing atomic optical transitions with H{\sc \,i} 21cm
and molecular millimetre transitions may yield an order of magnitude
over the already highly sensitive optical results
\cite{cs95,dwbf98,mwf+00}. However, very few such constraints are
available due to the paucity of quasar absorption systems where 2 of
the 3 types of transition (optical atomic, H{\sc \,i} 21cm or
molecular millimetre) exist.
\end{enumerate}
It is this last point which is of interest to us: As well as providing
a comprehensive list of these objects for use by the astronomical
community in general (Section 2), this catalogue allows us to
shortlist those DLAs most likely to exhibit radio absorption lines in
order to further constrain the variation in fine-structure
constant. In the final sections we present the DLAs occulting radio-loud quasars
along with any radio absorption features published and outline our
future plans regarding the sample.

\section{Explanatory comments and references}

In Table 1, for the column density we note that \scite{csb01} argue
that when $N_{\rm HI}$ is estimated directly from the absorption line
equivalent width, the value is systematically underestimated when
compared to the estimate derived using Voigt profile fitting.  This
may be due to a biases associated with estimating the quasar continuum
due the presence of Lyman forest absorption lines.  The value of
$N_{\rm HI}$ quoted in the table in this paper are those values
reported in the original sources, and are not corrected using the
relation given in Corbelli, Salpeter \& Bandiera (2001).

Note also that the visual magnitude is obtained from the DLA
reference, {V\'{e}ron-Cetty} \& {V\'{e}ron} (2001) or, failing these, NED,
which gives an approximate value. For the radio flux densities,
$S_{0.4}$ is either the 0.33 GHz WENSS, 0.37 GHz Texas or 0.41 GHz MRC flux density (see
Section 2.2), and where both flux densities are available the 365 MHz
value is quoted. In the case of the Texas survey, ``X'' denotes that
the quasar was not detected at the flux density limit of 0.25 Jy
({Douglas} {\rm et~al.} 1992). $S_{1.4}$, \etc are the measured 1.4
GHz, \etc continuum flux densities in Jy and for $S_{higher}$, the
frequency in GHz is given in parenthesis. In the appropriate columns,
``X'' also designates if the object is not considered a 2.7 or
5.0 GHz radio source ({V\'{e}ron-Cetty} \& {V\'{e}ron} 2001) and where
the 22 and 37 GHz values are quoted as approximate refers to the average
value from 15 years of monitoring by {Ter\"{a}sranta} {\rm et~al.} (1998). 
In general, approximate values are used when the
flux density is obtained from more than one reference and the values
given do not exactly agree. Finally, ``--'' denotes that no
information could be found. 

\subsection{The DLA references}

The DLAs are compiled from
\scite{wbcj78}$^{1}$, \scite{wd79}$^{2}$, \scite{wbj81}$^{3}$,
\scite{sbps82}$^{4}$, \scite{tyt82}$^5$, \scite{bgw+84}$^6$,
\scite{wtsc86}$^7$, \scite{bcf87}$^{8}$, \scite{tyt87a}$^9$,
\scite{tyt87b}$^{10}$, \scite{lan88}$^{11}$, \scite{ssb89}$^{12}$,
\scite{twl+89}$^{13}$, \scite{lan91}$^{14}$, \scite{lwt+91}$^{15}$,
\scite{ssg91}$^{16}$, \scite{bcj92}$^{17}$, \scite{cp92}$^{18}$,
\scite{my92}$^{19}$, \scite{bbb+93}$^{20}$, \scite{lwtl93}$^{21}$,
\scite{tb93}$^{22}$, \scite{wkb93}$^{23}$, \scite{csc+94}$^{24}$,
\scite{wft+94}$^{25}$, \scite{lstm95}$^{26}$, \scite{lwt95}$^{27}$,
\scite{sbbd95}$^{28}$, \scite{wlfc95}$^{29}$, \scite{bbs+96}$^{30}$,
\scite{cld+96}$^{31}$, \scite{dob96}$^{32}$, \scite{ipmw96}$^{33}$,
\scite{lsb+96}$^{34}$, \scite{ptsl96}$^{35}$, \scite{scc+96}$^{36}$,
\scite{smih96}$^{37}$, \scite{gb97}$^{38}$, \scite{lwa+97}$^{39}$,
\scite{lbbd97}$^{40}$, \scite{lsb97}$^{41}$, \scite{vcfm97}$^{42}$,
\scite{ihc98}$^{43}$, \scite{jbb+98}$^{44}$, \scite{lvm98}$^{45}$,
\scite{sp98}$^{46}$, \scite{lr99}$^{47}$, \scite{drt+00}$^{48}$,
\scite{mbc+00}$^{49}$, \scite{pasi00}$^{50}$, \scite{pes+00}$^{51}$,
\scite{rt00}$^{52}$, \scite{sw00}$^{53}$, \scite{btg01}$^{54}$,
\scite{coh01}$^{55}$, \scite{epss01}$^{56}$, \scite{kc01a}$^{57}$,
\scite{psm+01}$^{58}$, \scite{trn+01}$^{59}$, the radio selected QSO
survey of \scite{eyh+02}$^{60}$ and finally the five new DLAs at $z>3$
occulting PSS quasars of \scite{pgw01}$^{61}$.

Note that several of the DLA citations do not give the background
quasar's coordinates, although we did manage to get some (but not all)
of these from the authors. For the remaining few we obtained the
coordinates from elsewhere (\eg NED, SIMBAD, \pcite{vv01}). However,
note that when checking the coordinates for the DLAs without published
magnitudes against optical images (Digitized Sky and APM
Surveys)\footnote{These can be found at {\it
http://archive.stsci.edu/dss/} and {\it
http://www.ast.cam.ac.uk/$\sim$apmcat/}, respectively.}, we found no
optical counterparts at the limiting APM magnitudes of 21.5 (North)
and 22.5 (South) for QSOs 0112--30, 0115--30, 1052+04, 1159+01 (Hazard \&
McMahon, unpublished) nor QSO 1338+101 (Hazard \& Sargent, unpublished)
and so we have no knowledge of where these coordinates originally came
from. Where we have failed to find optical coordinates for any of
these sources, ``{\it not published}'' is inserted into the table. We
have kept these sources in the catalogue for the sake of providing a
full comprehensive list and, for the sake of consistency, we do not include the
best radio positions as this could prove misleading.

\subsection{The radio references}

The radio parameters of the background quasar supplying the continuum
emission are compiled from \scite{wm71}$^{a}$, \scite{cops81}$^{b}$,
\scite{gc91}$^{c}$, \scite{vif+92}$^{d}$, \scite{wb92}$^{e}$,
\scite{lbm93}$^{f}$, \scite{zss+94}$^{g}$, \scite{hifh95}$^{h}$,
\scite{kim+95}$^{i}$, \scite{bla96}$^{j}$, \scite{fsp96}$^{k}$,
\scite{omc+96}$^{l}$, \scite{bb97}$^{m}$, \scite{dbo97}$^{n}$,
\scite{hdr97}$^{o}$, \scite{kdhr98}$^{p}$, \scite{ldp98}$^{q}$,
\scite{ttm+98}$^{r}$, \scite{bmm+99}$^{s}$, \scite{opm+99}$^{t}$,
\scite{wk99}$^{u}$, \scite{pkkw00}$^{v}$, \scite{sdp+00}$^{w}$,
\scite{tlv00}$^{x}$, \scite{bhb+01} (and references therein)$^{y}$,
\scite{cbr+01}$^{z}$ and current versions of the Parkes (\pcite{wo90},
incorporating the 408 MHz flux densities from the MRC catalogue;
\pcite{lcb91})$^{P}$, Green Bank \cite{wb92}$^{G}$, FIRST
\cite{bwh95}$^{F}$, Texas \cite{dbb+96}$^{T}$, PMN
\cite{gwbe94,wgbe94,gwbe95,wgh+96}$^{pmn}$, WENSS \cite{rtd+97}$^{we}$, NVSS
\cite{ccg+98}$^{N}$, \scite{knk+99}$^{K}$ and \scite{vv01}$^{V}$
catalogues, and most recently, the MAMBO survey of the highest
redshift PSS quasars \cite{ocb+01}$^{M}$. Finally, the results of
\scite{wbg+00}$^{W}$ give new DLAs as well as the FIRST radio
luminosities of the quasars illuminating these.

Note that no radio information for the quasars in Table 1 was found in
the 5 GHz catalogue of \scite{bwe91}, the optically quiet quasar
search of \scite{kfl+95}, the S5 radio source catalogue of
\scite{sk96a}, the 100 and 150 GHz observations of Southern
flat-spectrum sources \cite{bcb+97}, 
%the WENSS 0.33 GHz Northern survey \cite{rtd+97}, 
the GHz peaked sample of \scite{sod+98} nor the
flat-spectrum sources of \scite{tuwv01}.

%for original \textheight=22.6 cm, y placement of table is 25.0	
%\newpage %this should be move to fill gaps in paper

\begin{table}[hbt] 
\setlength{\unitlength}{1cm} 
\begin{picture}(20,20)
\put(17.8,21.2){\includegraphics{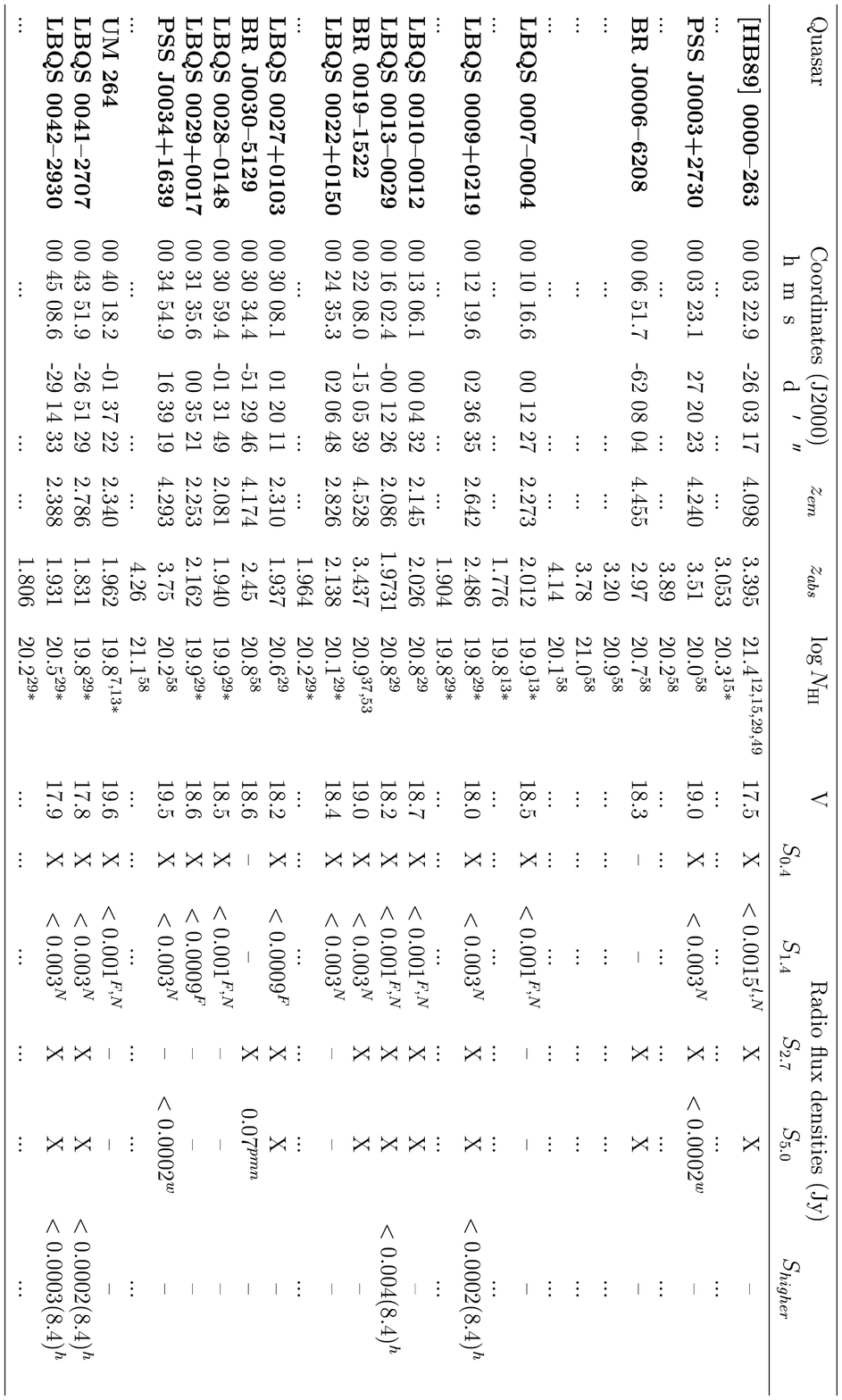}} 
%\put(17.8,20.5){\special{psfile=t1.ps hscale=100 vscale=100 angle=180}}
\end{picture}
\end{table}

\newpage
\begin{table}[hbt] 
\setlength{\unitlength}{1cm} 
\begin{picture}(20,20)
\put(17.8,17.7){\includegraphics{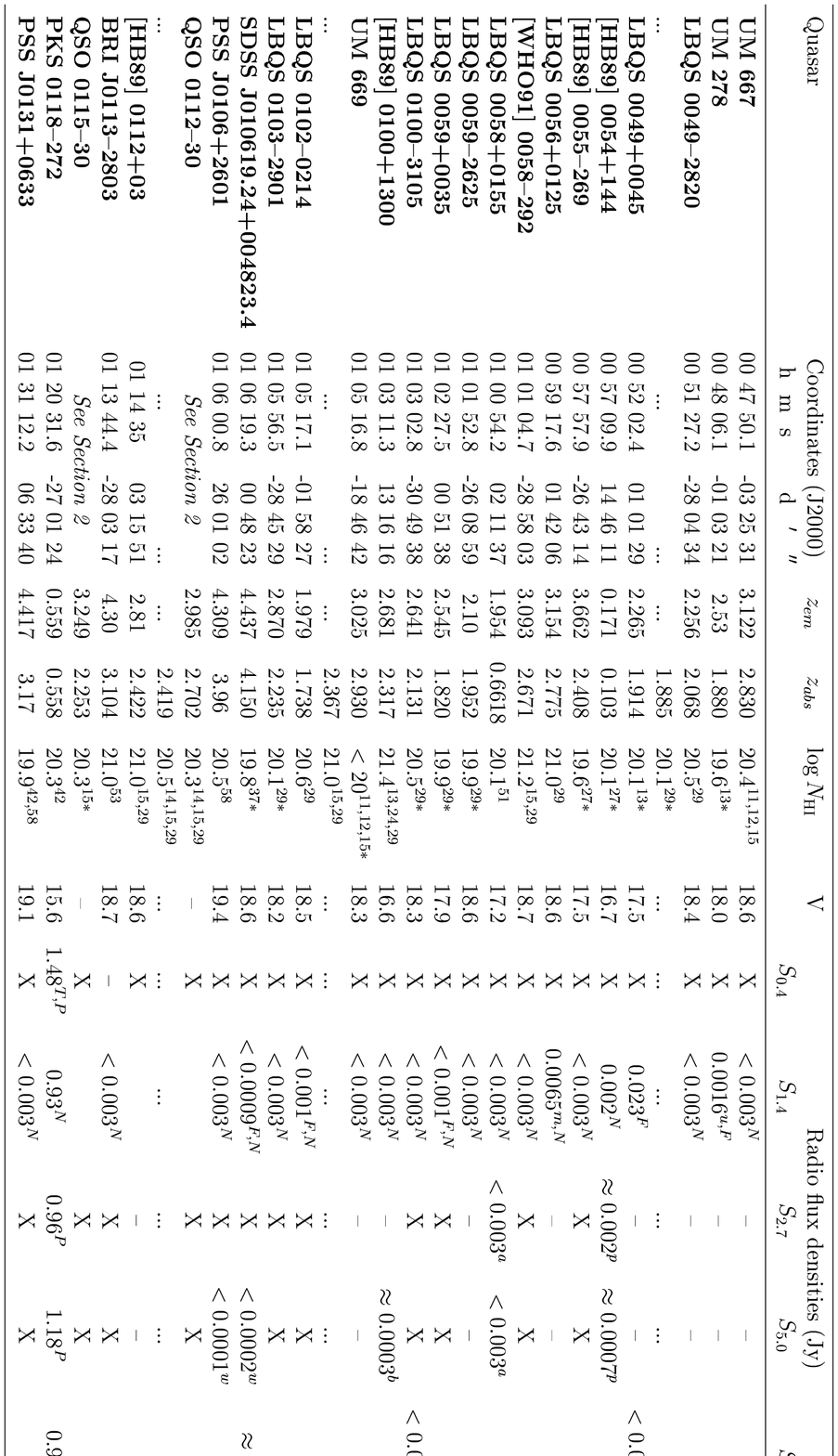}} 
\end{picture}
%\addtocounter{table}{-1} %next will be \addtocounter{table}{-2}  etc
\end{table}

\newpage %these are needed to get the unnumbered  pages in the right place
\thispagestyle{empty}
\begin{table}[hbt] 
\setlength{\unitlength}{1cm} 
\begin{picture}(20,20)
\put(17.8,17.2){\includegraphics{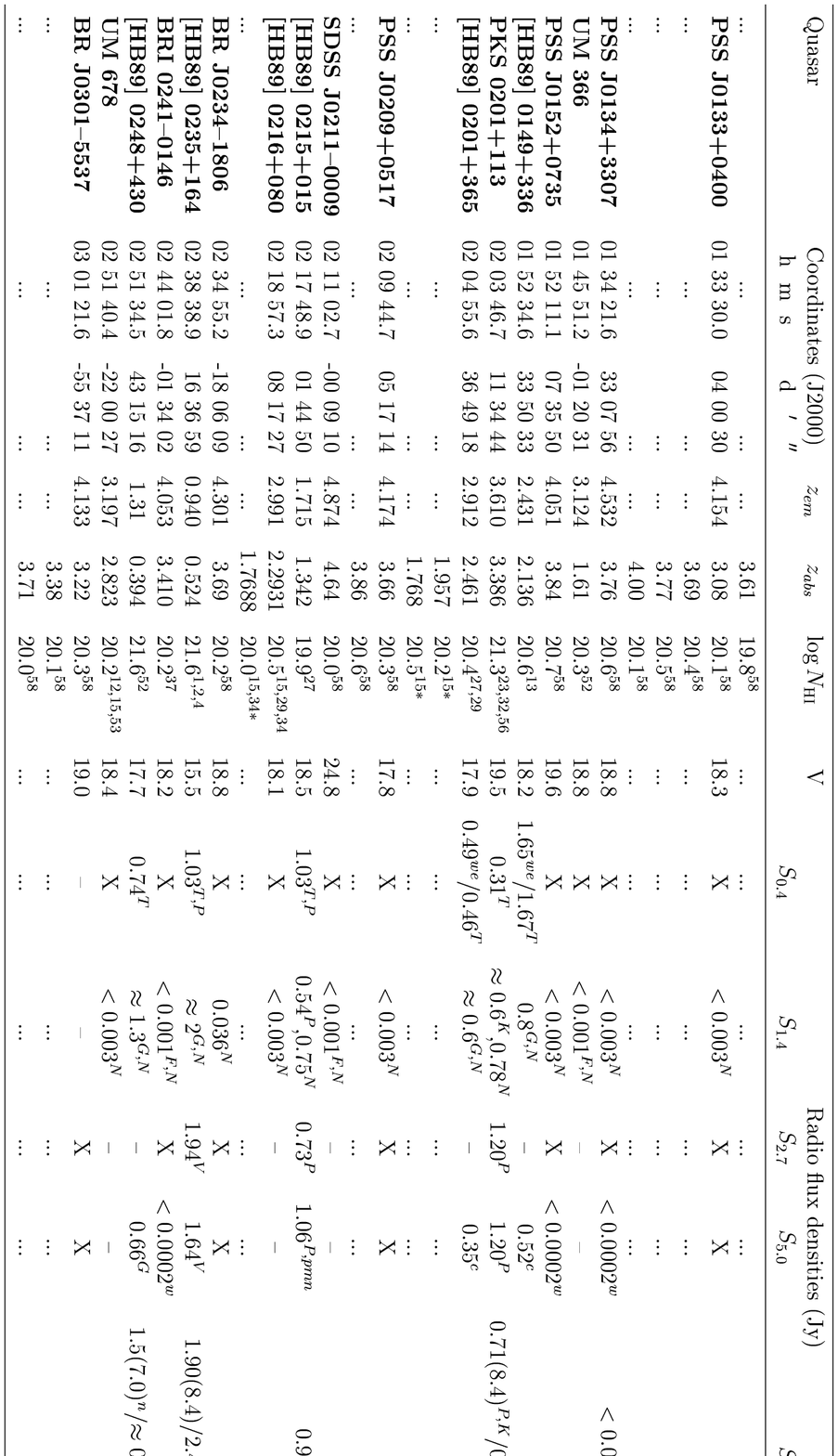}} 
\end{picture}
%\addtocounter{table}{-1} %next will be \addtocounter{table}{-2}  etc
\end{table}
\newpage
\begin{table}[hbt] 
\setlength{\unitlength}{1cm} 
\begin{picture}(20,20)
\put(17.8,17.7){\includegraphics{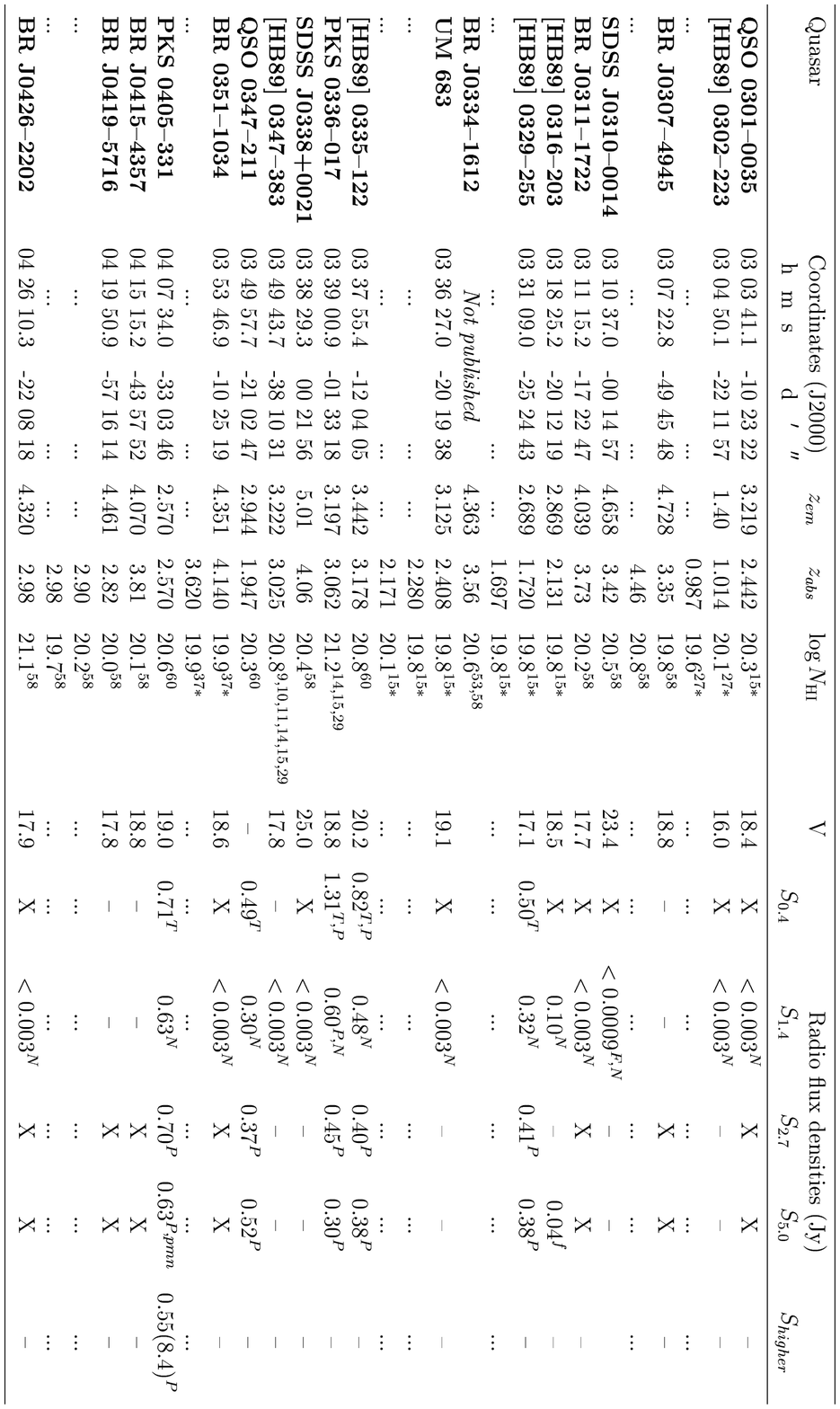}} 
\end{picture}
%\addtocounter{table}{-1} %next will be \addtocounter{table}{-2}  etc
\end{table}
\newpage   %these are needed to get the unnumbered  pages in the right place
\thispagestyle{empty}
\begin{table}[hbt] 
\setlength{\unitlength}{1cm} 
\begin{picture}(20,20)
\put(17.8,15.7){\includegraphics{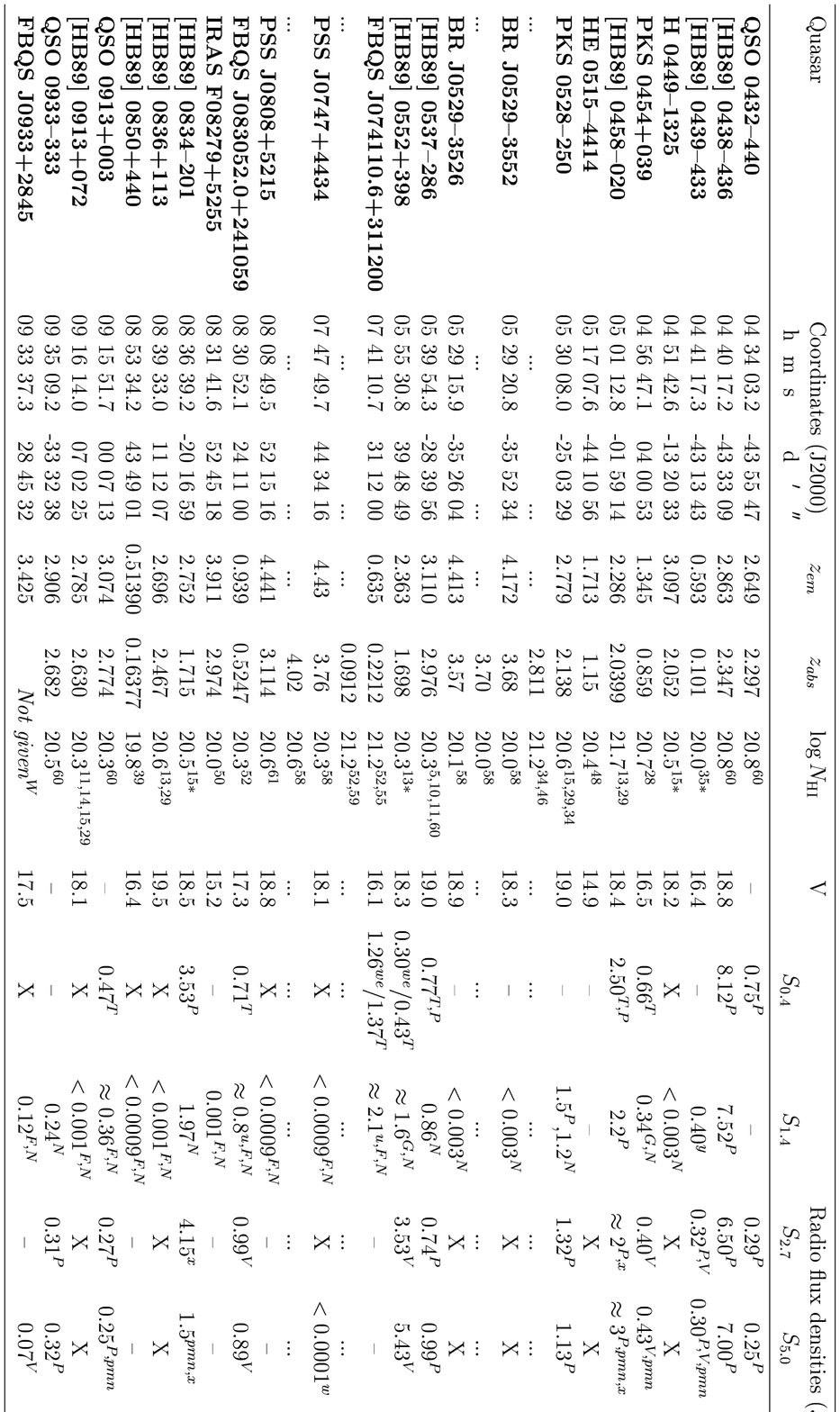}} 
\end{picture}
%\addtocounter{table}{-1} %next will be \addtocounter{table}{-2}  etc
%\caption{{\it continued}}
\end{table}
\newpage
\thispagestyle{empty}
\begin{table}[hbt] 
\setlength{\unitlength}{1cm} 
\begin{picture}(20,20)
\put(17.8,17.2){\includegraphics{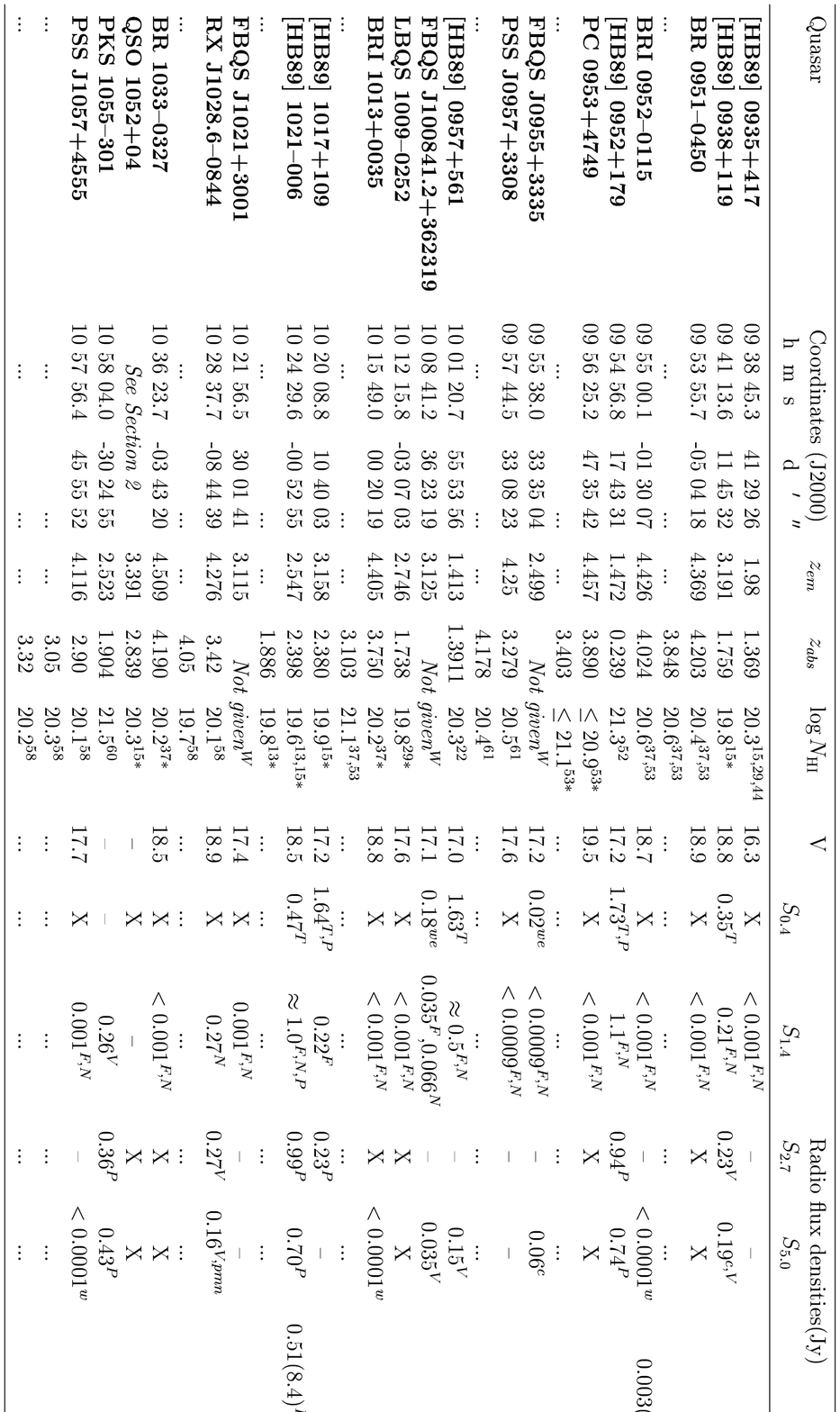}} 
\end{picture}
%\addtocounter{table}{-1} %next will be \addtocounter{table}{-2}  etc
%\caption{{\it continued}}
\end{table}
\newpage
\thispagestyle{empty}
\begin{table}[hbt] 
\setlength{\unitlength}{1cm} 
\begin{picture}(20,20)\
\put(17.8,17.2){\includegraphics{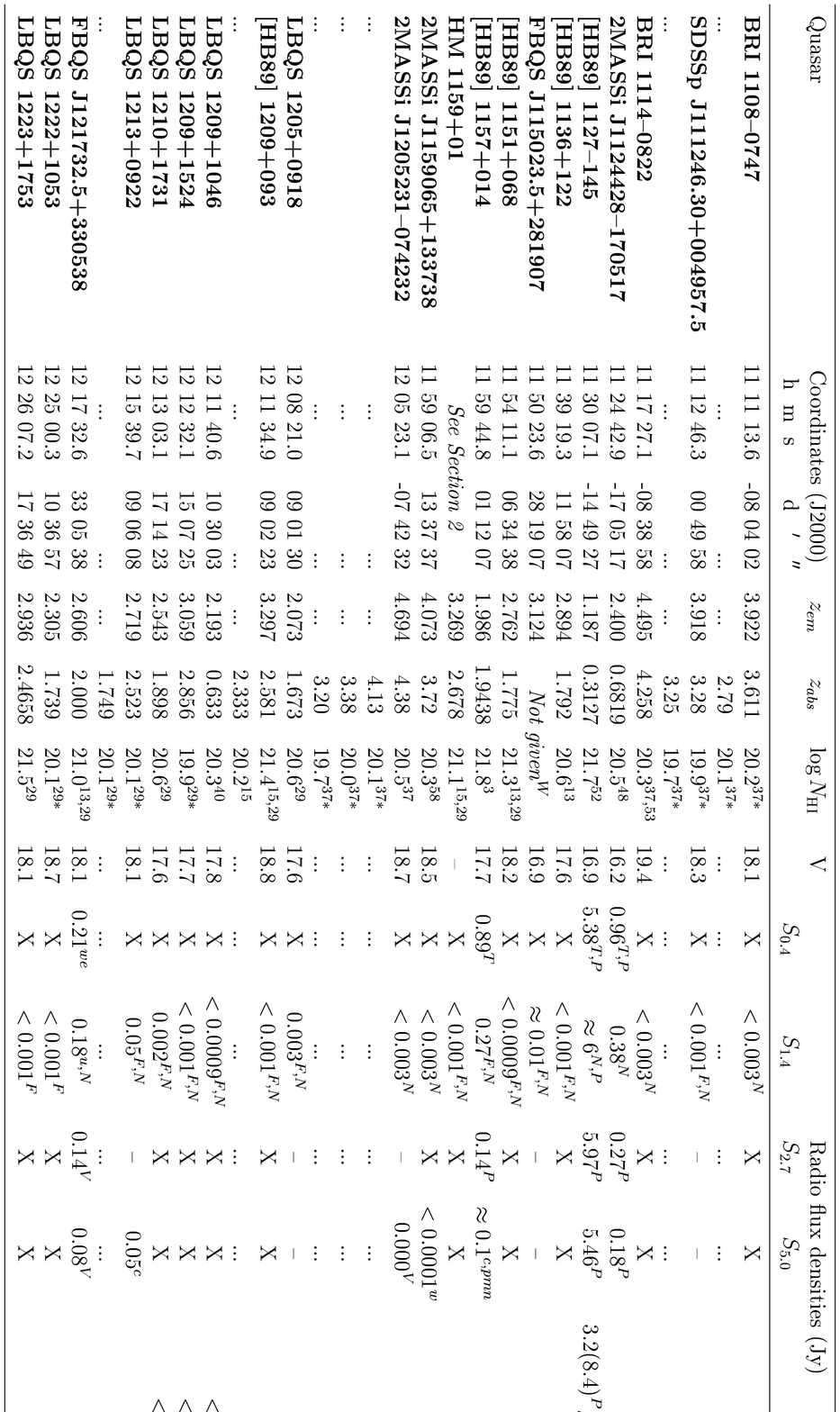}} 
\end{picture}
%\addtocounter{table}{-1} %next will be \addtocounter{table}{-2}  etc
%\caption{{\it continued}}
\end{table}
\newpage
\thispagestyle{empty}
\begin{table}[hbt] 
\setlength{\unitlength}{1cm} 
\begin{picture}(20,20)
\put(17.8,16.2){\includegraphics{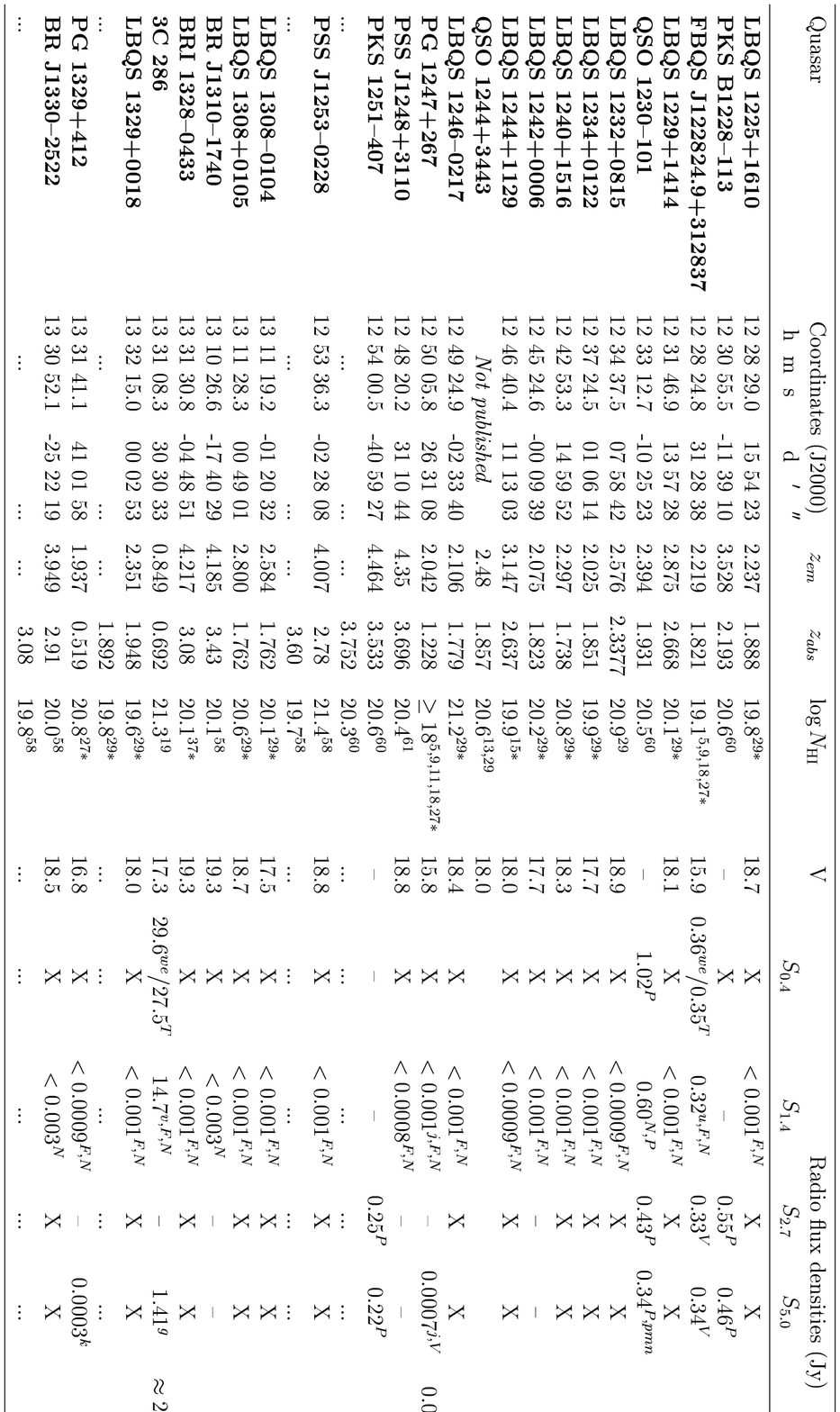}} 
\end{picture}
%\addtocounter{table}{-1} %next will be \addtocounter{table}{-2}  etc
%\caption{{\it continued}}
\end{table}
\newpage
\begin{table}[hbt] 
\setlength{\unitlength}{1cm} 
\begin{picture}(20,20)
\put(17.8,17.7){\includegraphics{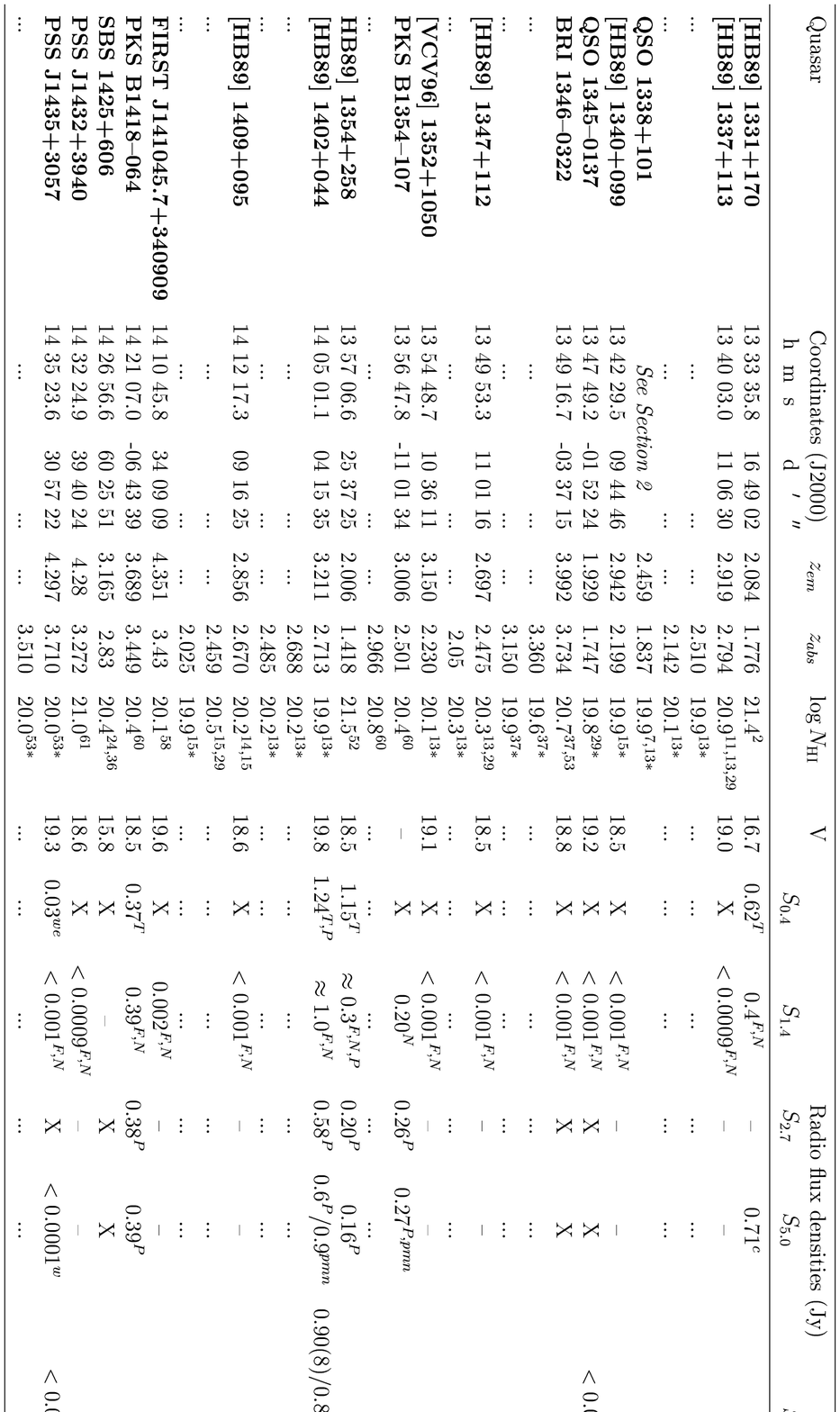}} 
\end{picture}
%\addtocounter{table}{-1} %next will be \addtocounter{table}{-2}  etc
%\caption{{\it continued}}
\end{table}
\newpage
\begin{table}[hbt] 
\setlength{\unitlength}{1cm} 
\begin{picture}(20,20)
\put(17.8,17.7){\includegraphics{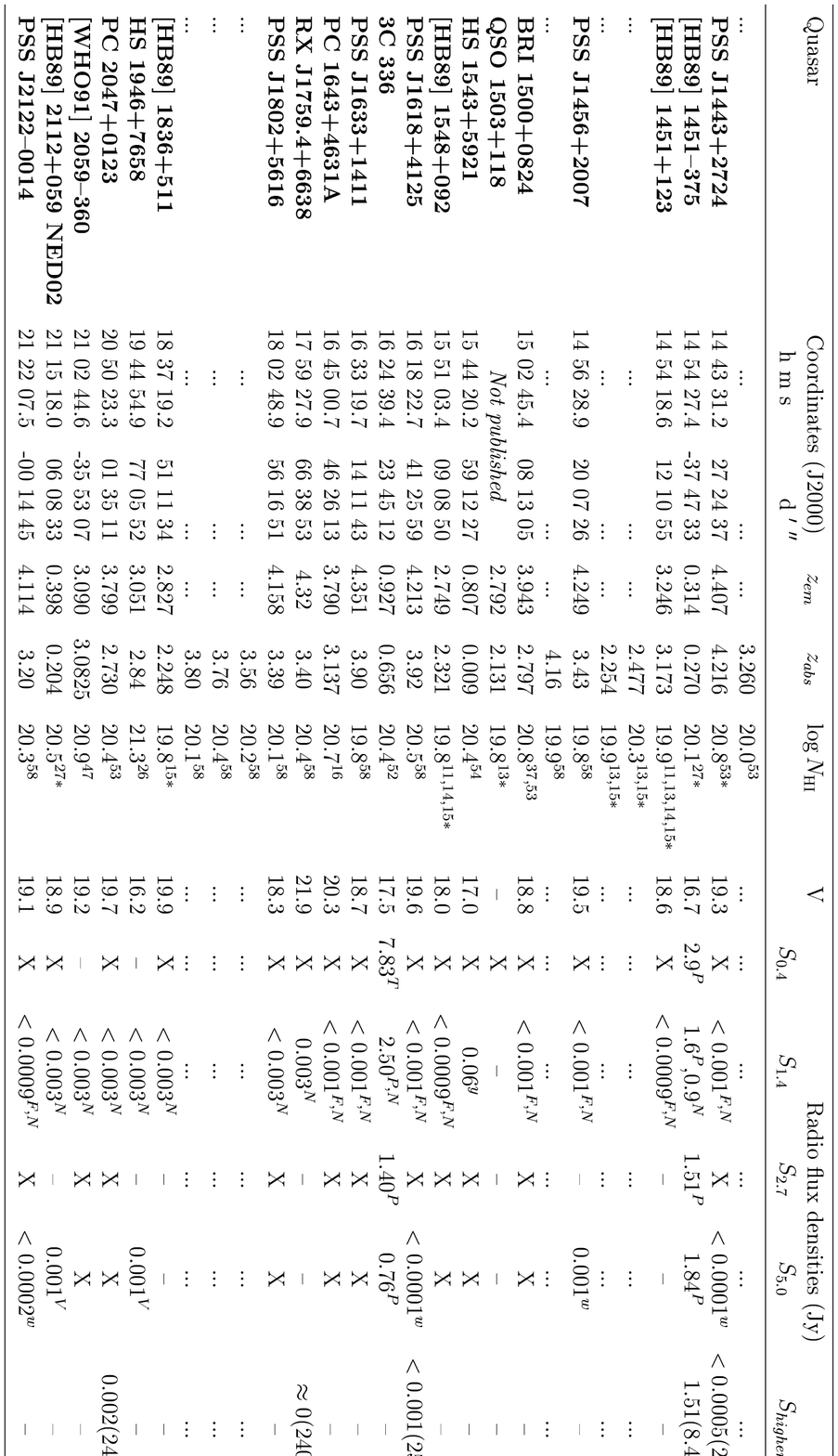}} 
\end{picture}
%\addtocounter{table}{-1} %next will be \addtocounter{table}{-2}  etc
%\caption{{\it continued}}
\end{table}
\newpage   %these are needed to get the unnumbered  pages in the right place
\thispagestyle{empty}
\begin{table}[hbt] 
\setlength{\unitlength}{1cm} 
\begin{picture}(20,20)
\put(17.8,16.2){\includegraphics{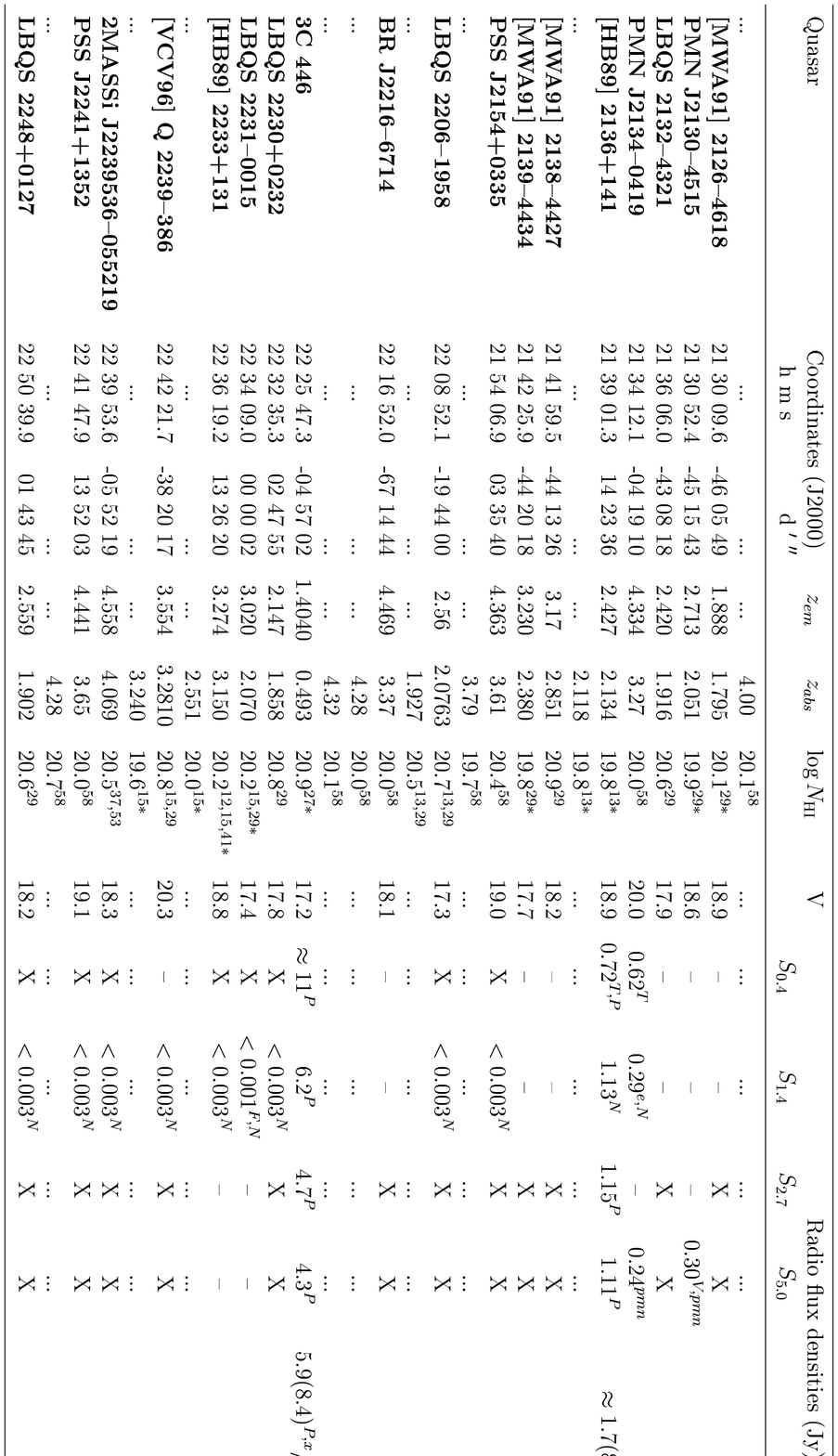}} 
\end{picture}
%\addtocounter{table}{-1} %next will be \addtocounter{table}{-2}  etc
%\caption{{\it continued}}
\end{table}
\newpage 
\begin{table}[hbt] %last (small) table landscape - doesn't work
\setlength{\unitlength}{1cm} 
\begin{picture}(20,20)
\put(17.8,17.7){\includegraphics{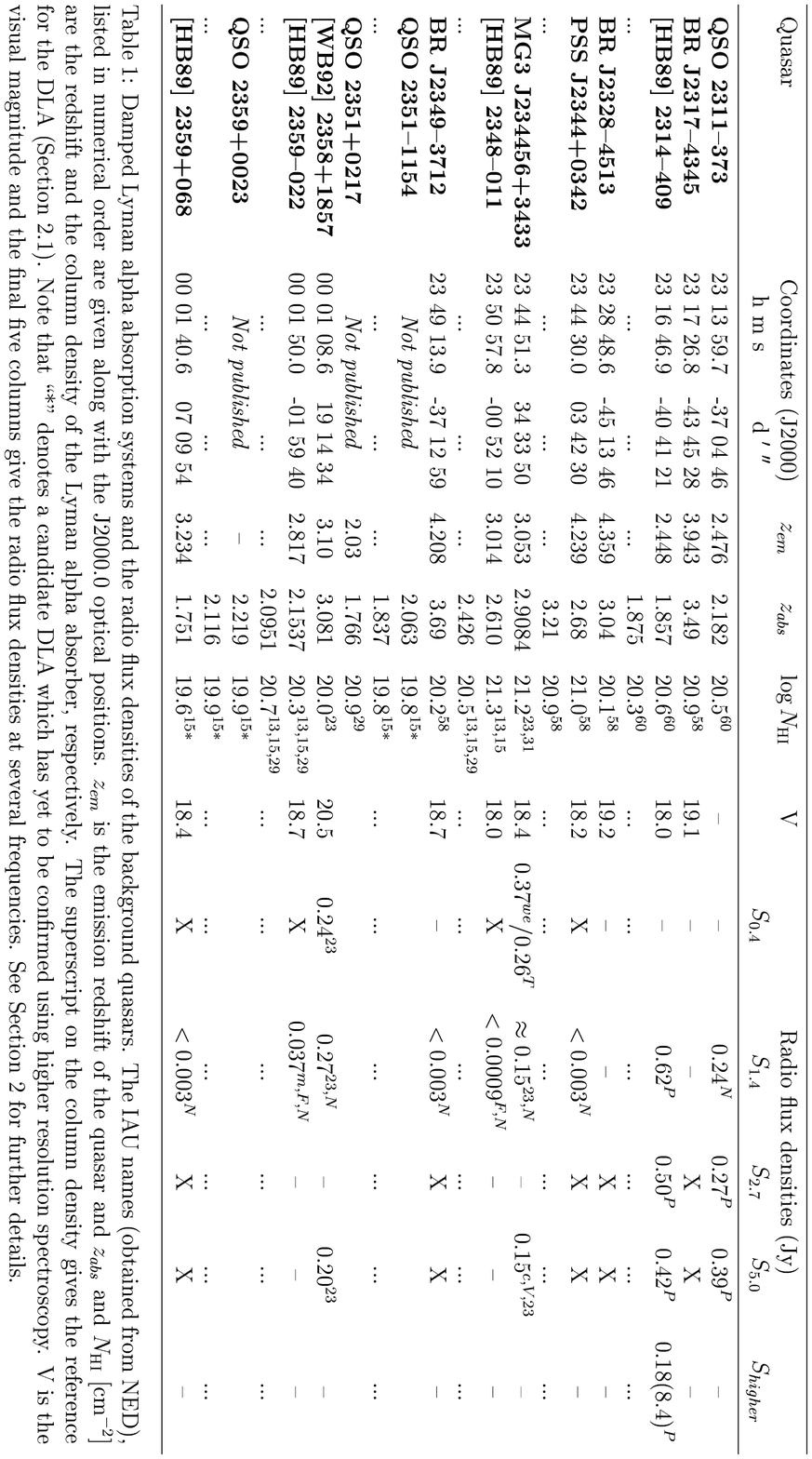}} 
\end{picture}
%\addtocounter{table}{-1} %next will be \addtocounter{table}{-2}  etc
%\caption{{\it continued}}
\end{table}

%\newpage

\clearpage

\section{Discussion}

As mentioned in the introduction, we have compiled this catalogue
since the comparison between optical and radio absorption lines can
provide a considerably more precise determination of
$\Delta\alpha/\alpha$: To a first approximation, the ratio of two
optical transition frequencies used in the many-multiplet method
\cite{dfw98b,wfc+98} is $\frac{\omega_1}{\omega_2} \propto
1+0.1\alpha^2$. However, the ratio of the hyperfine neutral hydrogen
(21\,cm) to an optical resonance transition frequency is directly
proportional to $\alpha^2$ \ie about 10 times larger. Thus, a
substantial improvement in the determination of any variation of
$\alpha$ could be made by obtaining further statistics from optical
{\it and} 21 cm lines in cosmological absorbers. The limit on the
variation of $\alpha$ can be obtained by the comparison of the \HI~ 21
cm line with any other optical or radio line (Section 3.2). However,
by using redshifted 21cm \HI~ together with $\alpha$-sensitive species
such as iron, zinc, chromium and nickel \cite{dfw98b}, frequently seen
in DLAs, we simultaneously maximise sensitivity and take advantage of
the different signs of the frequency shifts due to $\alpha$ variation
to help minimise systematic effects \cite{mwf+01b,wmf+01}.

A new systematic effect which applies to tests for
$\Delta\alpha/\alpha$ involving a \HI~ \& optical comparison involves the
possible different spatial characteristics of the radio and optical
quasar emission.  Large differences can result in the radio and
optical light probing slightly different lines-of-sight.  However, we
note that there are examples where the radio and optical emission is
known to coincide spatially, and those cases are clearly of particular
interest (Section 3.1). In order to minimise the spatial segregation problem, the
most reliable tests will come from comparing \HI~ lines with neutral
atomic or molecular species, or singly ionised species where the
ionisation potential is smaller than that for neutral hydrogen.

\subsection{Radio-loud quasars illuminating DLAs}
%\label{radio}

Of the known radio-loud ($S_{\rm radio}\gapp0.1$ Jy) systems, we
summarise the current state of searches for atomic and molecular
hydrogen (Section 3.2) absorption features. Note that with regard
to the spatial distribution of the optical and radio emission, from
the NVSS catalogue \cite{ccg+98}, {\it unless otherwise stated, the
1.4 GHz emission extends to a radius of $\approx1'$ and the
peak emission coincides with the given optical position (Table 1).}

\noindent {\bf PKS 0118--272:} A BL Lac object where \scite{kc01a} failed
to detect \HI~ absorption at $z=0.5579$.

\noindent {\bf [HB89] 0149+336:} A gravitational lens candidate for which we could find no reference to radio absorption features. 

\noindent{\bf PKS 0201+113:} A gravitational lens
where \scite{dob96,bbw97} have detected \HI~ absorption at
$z=3.388$.

\noindent{\bf [HB89] 0201+365:} No reference to radio absorption features found.

\noindent{\bf [HB89] 0215+015:} A BL Lac object where \scite{bw83} failed
to detect \HI~ absorption. 

\noindent{\bf [HB89] 0235+164:} A BL Lac object where \HI~ absorption at $z=0.524$ has been detected (\pcite{wbd82}; \pcite{bw83}). \scite{drrw92,wc95} failed to detect CO at the absorption redshift.

\noindent{\bf [HB89] 0248+430:} \scite{lb01} have detected the \HI~ absorption at the DLA redshift. 

\noindent {\bf [HB89] 0329--255:} No reference to radio absorption features found.

\noindent {\bf  [HB89] 0335--122:} No \HI~ absorption detected \cite{kc02}.

\noindent {\bf PKS 0336--017:} \scite{cld+96,kc02} failed to detect
\HI~ absorption at $z=3.0619$.

\noindent {\bf QSO 0347--211:} No reference to radio absorption features found. 

\noindent {\bf PKS 0405--331:} As above.

\noindent {\bf QSO 0432--440:} As above. {\it No NVSS data available.}

\noindent {\bf [HB89] 0438--436:} \scite{dcw96} failed to detect CO in the torus of this AGN ($z=2.852$). {\it No NVSS data available.}

\noindent {\bf [HB89] 0439--4319:} Tentative \HI~ absorption detected by
\scite{kcsp01} in this low redshift source. {\it No NVSS data available.}

\noindent {\bf PKS 0454+039:} No \HI~ \cite{bw83} nor H$_2$ \cite{gb99} absorption has been detected. {\it No optical/radio offset, but there is a second $30''$
radius radio source centered at 5 s to the West.}

\noindent {\bf [HB89] 0458--020:} In this blazar, \scite{wbt+85,bwl+89} have detected \HI~ absorption at $z=2.03945$. No H$_2$ nor CO (\ie molecular) absorption has been detected \cite{wc94a,gb99}.

\noindent {\bf PKS 0528--250:} \scite{cld+96} failed to detect \HI~
absorption at $z=2.8110$, although H$_2$ absorption in this DLA
\cite{fcb88,sp98,gb99} and CO emission in the $z=2.14$ DLA
\cite{bv93} have been detected. Note that no H$_2$ or CO absorption in either DLA was
detected by \scite{wc94a,lsb99}. 

\noindent {\bf [HB89] 0537--286:} No reference to radio absorption features found. 

\noindent {\bf [HB89] 0552+398:} Although Galactic H{\sc \,i} \cite{dkvh83} and HCO$^+$ \cite{ll96} absorption have been observed towards this quasar, no reference to absorption at the
DLA (or any cosmological) redshift could be found.

\noindent {\bf J074110.6+311200:} In this optically variable quasar, \scite{lsb+98,kgc01} have detected \HI~ absorption at $z=0.2212$.

\noindent {\bf FBQS J083052.0+241059:} In this blazar, \scite{kc01a} have detected \HI~ absorption at $z=0.5247$.

\noindent {\bf IRAS F08279+5255:} A gravitational lens in which \scite{cmo99} have detected CO 4$\rr$3 emission at $z=3.911$, the redshift of the source. {\it
There is a weak central radio source at optical position with two stronger diagonally
opposing sources near} 08h31m50s/52d$43'30''$ {\it and} 08h31m25s/52d$46'30''$. 

\noindent {\bf [HB89] 0834--201:} No reference to radio absorption features found for this blazar.

\noindent {\bf QSO 0913+003:} No reference to radio absorption features found. 

\noindent {\bf QSO 0933--333:} As above. {\it Offset from optical position at} 
09h35m08.6s/-33d$32'34''$.

\noindent {\bf [HB89] 0938+119:} No reference to radio absorption features found. {\it No offset but there is a second source to the South East near}
09h41m20.5s/11d$45''00'$.

\noindent {\bf [HB89] 0952+179:} \scite{kc01a} have detected \HI~ absorption at $z=0.2378$. 

\noindent {\bf [HB89] 0957+561:} A gravitational lens
where no \HI~ absorption has been detected \cite{kc02}.

\noindent {\bf [HB89] 1017+109:} No reference to radio absorption features found. {\it Radio position offset $\approx20''$ to the West
of the optical centre.}

\noindent {\bf [HB89] 1021-006:} No reference to radio absorption features found for this optically variable quasar.

\noindent {\bf RX J1028.6-0844:} No \HI~ absorption detected \cite{kc02}.

\noindent {\bf PKS 1055--301:} No reference to radio absorption features found. {\it Radio position offset $\approx1'$ to the West
of the optical centre.}

\noindent {\bf 2MASSi J1124428--170517 :}  No reference to radio absorption features found. {\it Offset slightly from optical position at} 
11h24m41.5s/-17d$05'10''$.

\noindent {\bf [HB89] 1127--145:} \scite{lsb+98,ck00} have detected variable \HI~ absorption at $z=0.3127$ towards this blazar.

\noindent {\bf [HB89] 1157+014:} \scite{wbj81,bw83} have detected  \HI~ absorption at $z=1.94362$. 

\noindent {\bf LBQS 1213+0922:} No reference to radio absorption features found.

\noindent {\bf FBQS J121732.5+330538 :} \scite{wc94a} failed to detect CO absorption at $z=1.9984$.

\noindent {\bf FBQS J122824.9+312837:} \scite{bw83} failed to detect \HI~ absorption at $z=1.7945$.

\noindent {\bf PKS B1228--113:} No reference to radio absorption features found. 

\noindent {\bf QSO 1230--101:} As above.

\noindent {\bf PKS 1251--407:} As above. {\it No NVSS data available.}

\noindent {\bf 3C 286:} \HI~ at $z=0.69215$ by \scite{br73} but no H$_2$ absorption has yet been detected \cite{gb99}.

\noindent {\bf [HB89] 1331+170:} A blazar where \scite{wd79,bw83} have detected \HI~ absorption at $z=1.7764$, but \scite{lsb99} failed to detect CO absorption.

\noindent {\bf PKS B1354--107:}. No \HI~ absorption
detected \cite{kc02} in the $z_{abs}=2.966$ DLA. {\it Radio position
offset $\approx15''$ to the West of the optical centre.}

\noindent {\bf [HB89] 1354+258:} No reference to radio absorption features found.

\noindent {\bf [HB89] 1402+044:} No reference to radio absorption features found for this BL Lac. 

\noindent {\bf PKS B1418--064:} No reference to radio absorption features found. 

\noindent {\bf [HB89] 1451--375:} \scite{ck00} failed to detect \HI~ absorption in this HST source.

\noindent {\bf 3C 336:} No reference to radio absorption features found for this optically variable quasar.

\noindent {\bf PMN J2130--4515:} No reference to radio absorption features found. {\it No NVSS data available.}

\noindent {\bf PMN J2134--0419:} No reference to radio absorption features found.

\noindent {\bf [HB89] 2136+141:} No reference to radio absorption features found.

\noindent {\bf 3C 446:} A blazar not detected in \HI~ absorption at the DLA \cite{ck00} nor CO absorption at the quasar redshift \cite{dcw96}.

\noindent {\bf QSO 2311--373:} No reference to radio absorption features found. 

\noindent {\bf [HB89] 2314--409:} As above. {\it No NVSS data available.}

\noindent {\bf MG3 J234456+3433:} \scite{cld+96,kc02} failed to detect \HI~ absorption at $z=2.9084.$ 

\noindent Finally, note that \HI~ absorption has been observed in the inferred (from
metal lines) DLAs
{\bf 3 C196}, {\bf LBQS 1229--0207} \cite{wlfc95} and {\bf [HB89] 1243--072} \cite{lb01}.

\subsection{Searching for new radio absorbers}
If we summarise the current \HI~ absorption results for the DLAs (Table 2),
we see that although many of the positive results have very high
column densities, this does not appear to be a prerequisite for \HI~
absorption (\ie FBQS J083052.0+241059). Perhaps also of relevance
is the spectral energy distributions (SEDs): Note that all
of the GHz peaked sources have high column densities and have
all been detected in \HI. Of the two inverted SEDs, one DLA
has a high column density whereas the other is relatively low and
\begin{table}
\begin{center}
\begin{tabular}{l r c c c l}
\hline\noalign{\smallskip}
Quasar & $\tau$ & $\log N_{\rm HI}$  & $S$ & S.I. & Notes\\
\noalign{\smallskip}
\hline\noalign{\smallskip}
PKS 0118--272 &  $<0.007$ & 20.3 & 1.2 & 0.1 & \\
PKS 0201+113 & 0.09,0.04 & 21.3 & 0.3 & -- & GPS (2.6)\\ 
 ${\rm[HB89]}$  0215+015 & $<0.04$ & 19.9 & 0.9 &-- &  See caption \\
 ${\rm[HB89]}$  0235+164 & 0.05-0.5 & 21.6 & 1.8 & -0.2 & Inverted \\
 ${\rm[HB89]}$  0248+430 & 0.20 & 21.6 & 1.2 & -- & GPS (2.5)\\ 
 ${\rm[HB89]}$  0335--122 & $<0.008$ & 20.8 & 0.8& 0.3 & \\%straightforward [HB89] screws up table
PKS 0336--017 &$<0.005$  & 21.2 & 1.3 & 0.6 & \\ 
 ${\rm[HB89]}$  0439--4319 & $<0.007$ & 20.0 & 0.4 & 0.2 & \\
PKS 0454+039 & $<0.01$& 20.7 & 0.4 & -- &  See caption \\
${\rm[HB89]}$  0458--020  & 0.3 & 21.7 & 2.5 & 0.3 & \\
PKS 0528--250 & $<0.2$ & 21.2 & 1.9 & 0.5 & For $z_{abs}=2.811$ DLA \\
J074110.6+311200 & 0.07 & 21.2 & 1.9 & -- & GPS (2.9)\\
 & & & &  &$z_{abs}=0.221$ DLA \\ 
FBQS J083052.0+241059 & 0.007 & 20.3 & 0.8 & -0.2 & Inverted\\
${\rm[HB89]}$  0952+179 &0.013 & 21.3 & 1.2 & 0.3 & \\
${\rm[HB89]}$  0957+561 & $<0.004$ & 20.3 & 0.9 & 1.3 & \\
RX J1028.6-0844  &  $<0.03$ & 20.1 & 1.7 & 0.9 & $z_{abs}=3.42$ DLA \\
${\rm[HB89]}$  1127--145 & 0.06 & 21.7 &6.2 &-- & GPS (1.4)\\ 
${\rm[HB89]}$  1157+014 & 0.05 & 21.8 & 1.0 & 0.8 & \\
FBQS J122824.9+312837 & $<0.05$ & 19.1 & 0.3 & 0 & \\
3C 286 & 0.11 & 21.3 & 19.0 & 0.6 & \\
${\rm[HB89]}$  1331+170 & 0.020 & 21.4 & 0.6 & -- &  See caption \\
PKS B1354--107 & $<0.05$ &20.8 & 0.2 & 0 & $z_{abs}=2.996$ DLA \\
${\rm[HB89]}$  1451--375 & $<0.006$ & 20.1 & 1.8 & 0.2 & \\
3C 446 &$<0.02$  & 20.9 & 7.4 & 0.5 & \\
MG3 J234456+3433  &$<0.04$  & 21.2 & 0.3 & 0.2 & \\
\noalign{\smallskip}
\hline
\end{tabular}
\label{sum}
\addtocounter{table}{1} %next will be \addtocounter{table}{-2}  etc
\caption{The radio-loud DLAs in which \HI~ absorption has been
searched for.  $\tau$ is the optical depth of the \HI~ line, with $3\sigma$
upper limits quoted, as given by the references in Section 3.1. For PKS
0201+113 the values are from \protect\scite{dob96} and
\protect\scite{bbw97}, respectively. PKS 0336--017 and MG3 J234456+3433
these are \protect\scite{kc02} results; \protect\scite{cld+96} obtained
$\tau<0.02$ and 0.1, respectively. $S$ is the approximate flux density in
Janskys at $z_{abs}$ and S.I. is the spectral index (both are estimated
from the flux density values in Table 1 and $S\propto\nu^{-{\rm
S.I.}}$). In the last column, GPS designates a GHz peaked source with the
approximate turnover frequency given in parenthesis.  In the case of the
``U-shaped'' SEDs: ${\rm[HB89]}$ 0215+015 is known to exhibit radio
outbursts (\eg \protect\pcite{lo85}) and so the flux densities quoted will
be variable. For PKS 0454+039 and ${\rm[HB89]}$ 1331+170, these could be
due to an anomalous flux density measurement and both are considered flat
spectrum sources (\eg \protect\cite{wgbb84,msm+97}).}
\end{center}
\end{table}
both of the flat SED detections have high column densities. Finally, the
two steep spectrum quasars which illuminate DLAs detected in \HI~
absorption ([HB89] 1157+014 and 3C 286) also have high column densities.

Because of the relation between turnover frequency and source size
\cite{ffs+90,ob97}, we may expect a higher \HI~ absorption detection
rate from flat and inverted SED sources, since these result from
similar optical and radio lines-of-sight.  However, as it stands, the
statistics are too small (Table 2) and so in order to maximise our
sample, it appears that the way to proceed is an unbiased search for
H{\sc\,i} in the DLAs occulting the remaining radio-loud quasars.

As mentioned in Section 1, as well as optical and \HI~ comparisons,
the inter-comparison of atomic and molecular lines will also give a
ten-fold increase in accuracy for $\Delta\alpha/\alpha$: Due to its
zero dipole moment and small moment of inertia, molecular hydrogen
cannot be directly observed at radio frequencies\footnote{In the case
of $z>1.8$ sources, however, the ultra-violet lines of H$_2$ are
redshifted into the optical window, making molecular hydrogen readily
observable at these frequencies. As well as for PKS 0528--250 (Section
3.1) molecular hydrogen has also been detected in the DLAs occulting
the radio-quiet quasars [HB89] 0000-263 \cite{lmc+00a}, LBQS 0013--0029
\cite{gb97,psl02}, [HB89] 0347--383 \cite{lddm01}, LBQS 1232+0815
\cite{gb97,spl00} and the inferred \cite{wlfc95} DLA [HB89] 0551--366
\cite{lsp02}.} and so it is the usual practice infer the presence of
this from the millimetre rotational lines of such molecules as CO. In
order to also take advantage of this, we have applied for time to
search
%\newpage
%\clearpage
for molecular absorption lines in the DLAs occulting mm-loud
quasars with the IRAM 30 metre and Swedish ESO Sub-millimetre
telescopes. Recently (April 2002), we have been awarded time on the
Australia Telescope Compact Array in order to obtain 90 GHz flux
measurements for the whole radio-loud sample, as a means of selecting
new sources in which to search for millimetre absorption. The results
will be published in forthcoming papers.

Also, with regard to finding new systems in which there may be
absorption (in all three frequency regimes), we see that 13 of the
quasars are known to be BL Lac/optically variable/blazars\footnote{BL
Lacs and Optically Violent Variables are known collectively as
blazars. In these radio-loud active galactic nuclei the radio jet is
relativistically beamed close to the line-of-sight (\eg
\pcite{pet97}).}
and that 3 of the sources
are known gravitational lenses.  This may be of interest as of
 the four known high redshift millimetre (\ie
molecular) absorbers, two are BL Lac objects; 
B 0218+357 \cite{wc95} and PKS 1413+135 \cite{wc94}\footnote{The former,
as well as PKS 1830--211, is also a gravitational lens \cite{wc96}.}.
This may suggest several
strategies for finding similar new absorbers \cite{sr99} which
could prove useful in appending to this catalogue.

\section{Summary}

We have performed an exhaustive search of the literature in order to
produce a list of all known damped Lyman alpha systems and their
associated radio properties. It is the 57 radio-loud
systems in which we are
interested as many of these have the potential to show \HI~ absorption in each DLA.
Of the sources searched, it is seen that several exhibit such absorption and
we are involved in an ongoing project to search for this in the remaining
systems. Not only will this give us radio lines for comparison with optical
data in order to constrain any temporal
variations in the fine structure constant, but we will have a significant sample
from which we could consider why some DLAs absorb in H{\sc\,i}
whereas others do not. For example, \scite{ck00}
suggest that low redshift DLAs may arise from a multitude of absorbers,
and hence do not have sufficient path length for \HI~ absorption,
while those of higher redshift are due to more compact systems. 
Finally, as well as finding new H{\sc \,i} absorbers, we hope that
this catalogue will prove useful to those using damped Lyman alpha systems
as part of their research. \\

\noindent {\bf Final note:} In order to retain its usefulness to the astronomical community,
we have now produced an on-line version of this catalogue which will be
continually updated. This is available from {\it http://www.phys.unsw.edu.au/$\sim$sjc/dla}.

%\clearpage
\section*{Acknowledgments}

We wish to thank the John Templeton Foundation for supporting this
work, the two anonymous referees and Barry Madore at NED for their
comments, Sandra Ricketts of the Anglo-Australian Observatory library
as well as Fredrik Rantakyr\"{o}, Sara Ellison, Ken Lanzetta and
especially Nissim Kanekar for giving us the preliminary results of his
\HI~ survey. This research has made use of the NASA/IPAC Extragalactic
Database (NED)\footnote{Operated by the Jet Propulsion Laboratory,
California Institute of Technology, under contract with the National
Aeronautics and Space Administration.} as well as the VizieR database
of CDS catalogues \cite{obm00}\footnote{{\it
http://vizir.u-strasbg.fr/local/cgi-bin/vizHelp?faq.htx}}.

%\section*{References}

%\newpage
%\bibliographystyle{/home/sjc/styles/mnras} 
%\bibliography{aa,ref} %A&A journal abbreviation style \bib\aa.bib

\end{document}